\documentclass[12pt]{article}

\usepackage[T1]{fontenc}
\usepackage{amsmath}
\usepackage{amsfonts}
\usepackage[english]{babel}

\usepackage{geometry}                
\usepackage[parfill]{parskip}    
\usepackage{graphicx}
\usepackage{amssymb}
\usepackage{epstopdf}
\usepackage[all]{xy}

\usepackage{amsthm} 
 
\theoremstyle{definition} 
 
\theoremstyle{axiom}

\theoremstyle{remark} 
 
\theoremstyle{plain}

\theoremstyle{plain}

\begin{document}

\title{Global warming in figures and the question of its treatment: some historical and epistemological views}
\author{Daniel Parrochia}
\date{University of Lyon (France)}
\maketitle

\textbf{Abstract.}
We first recall fundamentals of elementary climate physics: solar constant, radiative balance, greenhouse effect, astronomical parameters of the climate (theory of Milankovitch). Without disputing the analyzes of climatologists and the famous Keeling curve revealing in an indisputable way the increase in CO$_{2}$ in the atmosphere since the industrial revolution, we nevertheless insist on the main contributor to the greenhouse effect which is, as we know, water vapor. Faced with the difficulties that there will be in imposing zero-carbon policies everywhere in the world (and especially in developing countries), we show that it would perhaps be in our interest to act on soil drought, which amounts, in fact, to being interested in the clouds. The decrease in cloud cover, due to a lack of water fixation in the soil, in fact increases the general temperature and therefore the greenhouse effect. Acting on CO$_{2}$ will always have, in this context, much less effect than acting on water vapor, even indirectly. Despite the difficulty of making this action sustainable, due to the balance of atmospheric water vapor and the oceans, it would be in our interest not to neglect this path and also possibly increase forest cover for this purpose, given the problems of setting up zero-carbon policy on a global scale.
In desperation, one can also consider protecting the Earth with an artificial dust cloud.


\section{Solar constant}

The Sun is a black body with the average temperature $T = 5780^\circ$ K approximately. It emits in a half-free space a flux of surface energy quantitatively estimable by the so-called "Stefan-Boltzmann law", including the constant of the same name  $\sigma = 5.67037442 \times 10^{-8} kg\ s^{-3}\ K^{-4}$. We have:
\[
F_{\odot} = \sigma T_{\odot}^4 = 6.32.10^7 \; W.m^{-2}.
\]

At a distance of $R \simeq 15 \times 10^7$ kilometers, setting the conservation of energy radiated through space, we get:

\[
4 \pi R_{\odot}^2 F_{\odot} = 4 \pi R^2 F.
\]

From where:

\[
F = F_{\odot} \left( \frac{R_{\odot}}{R} \right)^2.
\]

And, with a stellar radius of 695,600 kilometers, $F$ is estimated to be:

\[
F=1362\; W.m^{-2}.
\]

N.B. The distances being approximate or rounded off, the numerical value of 1366 W, or even 1370 W, is sometimes found in the literature.

\section{Variations in solar radiation}

Since the formation of the solar system, about 4.7 billion years ago, the intensity of solar radiation has increased. At that time, it was worth only 70\% of its current value, and during the Carboniferous, around 300 million years ago, when the first dinosaurs appeared and tropical vegetation developed abundantly, the constant solar was about 2.5\% lower than today.

It has been shown\footnote{Historically, the first serious determination of the solar constant dates from 1838 and goes back to the French physicist Claude Pouillet (1790-1868) who estimates it at 1,228 W.m$^{-2}$. This value, however close to reality, was called into question in 1881 by Samuel Pierpont Langley who found the erroneous value of 2,140 W.m$^{-2}$ following an expedition to the summit of Mount Whitney (4,420 m). This value, which will be a reference for more than 20 years (see \cite{Duf}), will only be corrected with the putting into orbit of modern radiometers. In 1978, the HF radiometer on the Nimbus 7 satellite announced a value of 1,372 W.m$^{-2}$. This value will however be quickly refined to 1,367 W.m$^{-2}$ by ACRIM I on SMM. More recently, VIRGO on SoHO brought the latter down to 1,365.4 $\pm$ 1.3 W.m$^{-2}$ in 1998. Finally, since 2008, the value retained is equal to 1,360.8 $\pm$ 0.5 W.m$ ^{-2}$ (see \cite{Kop}).} that the solar constant, expressed as a percentage of its current value, can be described by the following equation (the time $t$ being expressed in billions of years since the appearance of the solar system):

\[
\left[ 1 + 0,4 \left(1 - \frac{t}{4,7} \right) \right]^{-1}.
\]

Thus, in 4.7 billion years, the Sun will be about 67\% more powerful than now, in terms of emitted radiation. Possible temporary variations lasting 10 million years (about every 300 million years) could explain the ice ages on Earth: for example, the Pleistocene (first epoch of the Quaternary) is an ice age, the preceding occurring between 300 and 700 million years ago.

But other terrestrial effects could be preponderant, such as the arrangement of the continents (around the poles) and the concentration of greenhouse gases.

The variation in the solar constant could be explained by the movement of the solar system around the Milky Way. The solar system rotates in the plane of the Galaxy in about 250 million years by oscillating. Every 33 million years, we cross the plane of the Galaxy; this is one of the hypotheses evoked to explain the significant climate changes, potentially at the origin of the massive disappearance of living species.

The solar constant also varies, of the order of 1 to 5 W.m$^{-2}$, on shorter time scales, from a few days to a few years, for example due to the presence or absence sunspots or solar activity.

Among the causes of global warming in the XX$^{th}$ and XXI$^{th}$ centuries, the Sun is estimated to be responsible for 10 to 12\%. It is likely that since 1750, when systematic meteorological records began (see \cite{Par1}), the Sun has increased the average temperature of the globe by 0.45$^\circ$C; the sunspot cycle was virtually absent during the Little Ice Age.

Nevertheless, in the long term, the astronomical position of the Earth in relation to the Sun is the main factor of natural variability in global temperature, through the solar "constant". The main cycles concern:

– Variations in the eccentricity of the Earth's orbit (current cycle of 100,000 years);

– Obliquity of the axis of the poles (current cycle of approximately 41,500 years);

– The precession of the equinoxes;

– Solar activity, which fluctuates according to an 11-year cycle, characterized by the number of sunspots.

The integration of all these factors and others within the framework of the astronomical theory of paleoclimates has given rise to various works in climatology (see \cite{Ber2}) and cyclostratigraphy (see \cite{Wee}; \cite{ Mey}).

\section{Earth's radiation balance}

The Earth's radiation balance\footnote{The idea of calculating a "radiation balance" does not seem to predate 1992 and the book by Peixoto and Oort (see \cite{Pei}, 91-130). We find in 2000 the notion of an "Earth energy balance" (see \cite{Kus}), and in 2005, a variant of it in the form of an "Earth energy imbalance" (see \cite{Han}). But the expression "Earth energy budget" appears only at the beginning of 2009 (see \cite{Tre}; \cite{Lin}).} is the algebraic sum of the energy received and lost by the Earth's climate system, namely the soil-atmosphere-ocean system.

  The Sun being a G2 type star, its emission spectrum extends from 0.2 to 4 micrometers, that is to say from ultraviolet to infrared passing through the visible.
 
  The sun radiates power P$_{total}$ in all directions in space, but at a distance $D$ this power is distributed over a fictitious sphere of radius $D$.
 
  The area of this sphere is S$_{Sphere} = 4 \pi D^2$.
 
  The surface solar power is therefore equal to $\frac{P_{total}}{S_{Sphere}} = \frac{P_{total}}{4 \pi D^2}$.
 
  The calculation of the solar power actually affecting the Earth supposes the existence of a virtual disk intercepting the solar radiation and of radius equal to the radius $R_{T}$ of the Earth. The surface of this disk is:
  \[
  S_{Disk} = \pi R^2_{T}.
  \]
   The power actually received by the Earth is therefore equal to the product of the surface solar power by the surface of the disc. It is therefore expressed as follows:
   \[
P_{Earth} = P_{Solar\ areal}\times \pi R^2_{T}.,
\]
Therefore, the power received by the earth is:
  \[
P_{Earth} = \frac{P_{total}}{4 \pi D^2} \times \pi R^2_{T} = \frac{P_{total}}{4 D^2} \times R^ 2_{T}.
\]
which can also be written:
\[
  P_{Earth} = \frac{R^2_{T}}{4 \times D^2} \times P_{Total}.
\]
The numerical application assumes the following values:
\[
R_{T}\ (radius\ of\ the\ Earth) = 6370 km = 6370 \times 10^3 m.
  \]
  \[
D\ (average\ distance\ Earth-Sun) = 149.5\ million\ km = 149.5 \times 10^9 m.
  \]
  So this gives us:
 
   \[
P_{Earth} = \frac{(6370 \times 10^3)^2}{4 \times (149.5 \times 10^9)^2} \times 3.86 \times 10^{26} = 1 .75 \times 10^{17} W.
  \]
  This power is distributed over the entire $S$ surface of the Earth, with:
  \[
  S = 4 \pi R^2_{T}.
  \]
  Which finally brings a local flux of the following power:
  \[
P_{local\ area} = \frac{P_{Earth}}{S} = \frac{1.75 \times 10^{17}}{4 \times \pi \times (6370 \times 10^3)^ 2} = 342W.
\]
 
 \section{Greenhouse effect and global warming}
 
  The Earth's atmosphere is mainly composed of the following gases:
 
  \begin{itemize}
  \item Nitrogen (N$_{2}$): about 78\%;
  \item Oxygen (O$_{2}$): about 21\%;
  \item Argon (Ar): about 1\%.
  \end{itemize}
 
  The rest is formed by rare gases, including CO$_{2}$ (carbon dioxide) which is measured, not in percentages, but partly by millions of molecules (ppm) with the rule: 1 ppm = 1/1000,000.
 
  Today, the rate of CO$_{2}$ in the atmosphere is around 400 ppm, or 0.04 \% of it, which may, a priori, seem very low. The concern stems from the fact that 1) CO$_{2}$ is presumed to be a greenhouse gas and 2) its quantity in the atmosphere has increased by 50\% since the advent of the industrial revolution in the 19th century. Is this concern justified?

\subsection{The origin of worry}

 It is good, before all, to recall in which circumstances this problem concerning the level of atmospheric CO$_{2}$ appeared and then amplified.
 
The idea that CO$_{2}$ emissions due to the combustion of fossil resources (coal, oil, gas, lignite, etc.) contributed to increasing terrestrial temperatures was put forward at the beginning of the XX$^{th}$ century by the Swedish scientist Svante Arrhenius, 1903 Nobel Prize in Chemistry. In a major book (see \cite{Arr}), he claimed that CO$_{2}$ contributed to the increase in temperatures by greenhouse effect, which, for him, was not a cause for concern but on the contrary presented a double advantage: 1. The atmospheric CO$_{2}$ stimulated the growth of plants; 2. It warmed the earth and made the human habitat more comfortable. As it was supposed, at that time, that the earth was rather cooling, the theses of Arrhenius fell into oblivion.They were only taken up again towards the end of the 1950s by the American oceanographer Roger Revelle (see \cite{Rev}). In 1957 exactly, the latter showed that most of the CO$_{2}$ of anthropogenic origin was absorbed by the oceans, and only marginally influenced global warming. He conjectured, however, that the level of atmospheric CO$_{2}$ would switch to a sudden increase capable of raising temperatures if, by chance, the industrial consumption of fossil fuels increased suddenly, to the point of no longer being able to be compensated by oceanic absorption. These catastrophic ideas of Revelle's, however, remained all the more a dead letter since, contrary to his predictions, the evolution of temperatures from 1950 to 1975 was rather oriented downwards (see \cite{Sch}) (it is possible that the presence aerosols of volcanic origin in the atmosphere has compensated for the influence of CO$_{2}$ emissions throughout this quarter of a century). Until 1987, in any case, the majority belief of scientists was that the world was heading towards a new glaciation. Thus in 1983, the British climatologist Hubert Lamb (see \cite{Lam}), in the first edition of his book on climate, predicted an annual drop in temperatures of -0.15$^{\circ}$C, until 2015 \footnote{If this prediction had turned out to be correct, from 1983 to 2023, we would have observed a colossal drop in temperatures of -6$^{\circ}$C in 40 years (-4.8$^{\circ}$C if we limit ourselves to the horizon of 2015). }. The prospect of this new ice age worried American experts so much that they set up several government task forces (see \cite{Deh}; \cite{Pos}; \cite{Ger}) to study these questions. This is how, paradoxically, the interest of the American authorities in global warming actually stems from the fear of a major cooling (which, fortunately, never happened). The IPCC\footnote{ Intergovernmental Panel on Climate Change.} (organization partly scientific, but partly also controlled by the States) as well as the cycle of the COPs\footnote{Conference of the Parties, the signatory states of the UNFCCC (United Nations Framework Convention on Climate Change) held in Rio in 1992.} are the consequences of all that). It was during the scorching drought of 1988, which mainly affected the USA, that James Hansen\footnote{Hansen has radicalized over the years, to the point of becoming an activist in the fight against global warming and being arrested in 2011 during a demonstration in the USA. His excessive and alarmist theses have, moreover, been criticized many times, notably by the astrophysicist Freeman Dyson, who accused Hansen "of transforming his science into ideology" (see \cite{Daw}).}, a NASA climatologist, succeeded in convincing the American Congress of the theses of Revelle\footnote{Hansen, studying the atmosphere of Venus , had been particularly struck by its very high temperatures (460$^{\circ}$C) which were entirely due, according to him, to a high concentration of CO$_{2}$ (the atmosphere of Venus indeed contains 250,000 times more CO$_{2}$ than that of the earth). He thus easily imagined that the increase in CO$_{2}$ levels in the Earth's atmosphere could lead, as for Venus, and all things considered, to excessive temperatures. The relevance of these theses jumped to the eyes of the deputies of the Congress during the episode of scorching drought of 1988, when the air conditioners of the meeting room failed to regulate the temperature.}. But other measures have come to support these theses.

\subsection{The Mann Curve Controversy}

Since 1988, therefore, and through the growing influence of the IPCC, an official discourse has developed to support, contrary to previous beliefs, but with equal certainty, that we will henceforth towards the heat, and no longer towards the cold. Over the course of the six Assessment Reports (from AR1 to AR6) – and the fifteen Special Reports (the last of which, SR15 was published in 2019), the IPCC has declined its alarmist forecasts for 2100: temperature rises of 0.6$^{\circ}$C and 1.1$^{\circ}$C compared to 2023 and a rise in sea level between 26 and 77cm in the first case, between 36cm and 87cm in the second. Less pessimistic forecasts due to Koolin, counting on a rise in sea level of 3mm per year, would still lead to a rise of about 24cm, even in the absence of limitation of anthropogenic CO$_{2}$ emissions (see \cite{Koo}). We remain close to the low forecasts of the IPCC. On the other hand, the forecasts become frankly debatable in the field of temperatures, where the models which confront each other reveal such divergences that they make the constructions which produce them little credible (variation of 1 to 3$^{\circ }$ or more).

Among the arguments advanced to support these theses, there is the idea that the end of the XX$^{th}$ century and the beginning of the XXI$^{th}$ would constitute the hottest years that the Earth has known for at least a thousand years, which then makes it possible to induce worrying extrapolations. The person in charge of this legend is an American scientist, Michael E. Mann (see \cite{Man}), at the origin of the eponymous curve, also known as "hockey stick". The graph of this curve, which describes the evolution of the temperatures of the planet during the last 1000 years, reveals a practically stable temperature from the year 1000 until 1950, followed by a brutal and catastrophic rise from 1950 to 2000. Unfortunately, this curve does not take into account variations that occurred a few centuries ago, such as the medieval optimum of the X$^{th}$-XIV$^{th}$ century or the "little ice age" of the XVII$^{th}$-XVIII$^{th}$ centuries in Europe. Hence a lively controversy with other researchers such as the Canadian climatologist Tim Ball (see Fig. 1):

\begin{figure}[h] 
\vspace{-1\baselineskip}
\hspace{5\baselineskip}
  \includegraphics[width=5in]{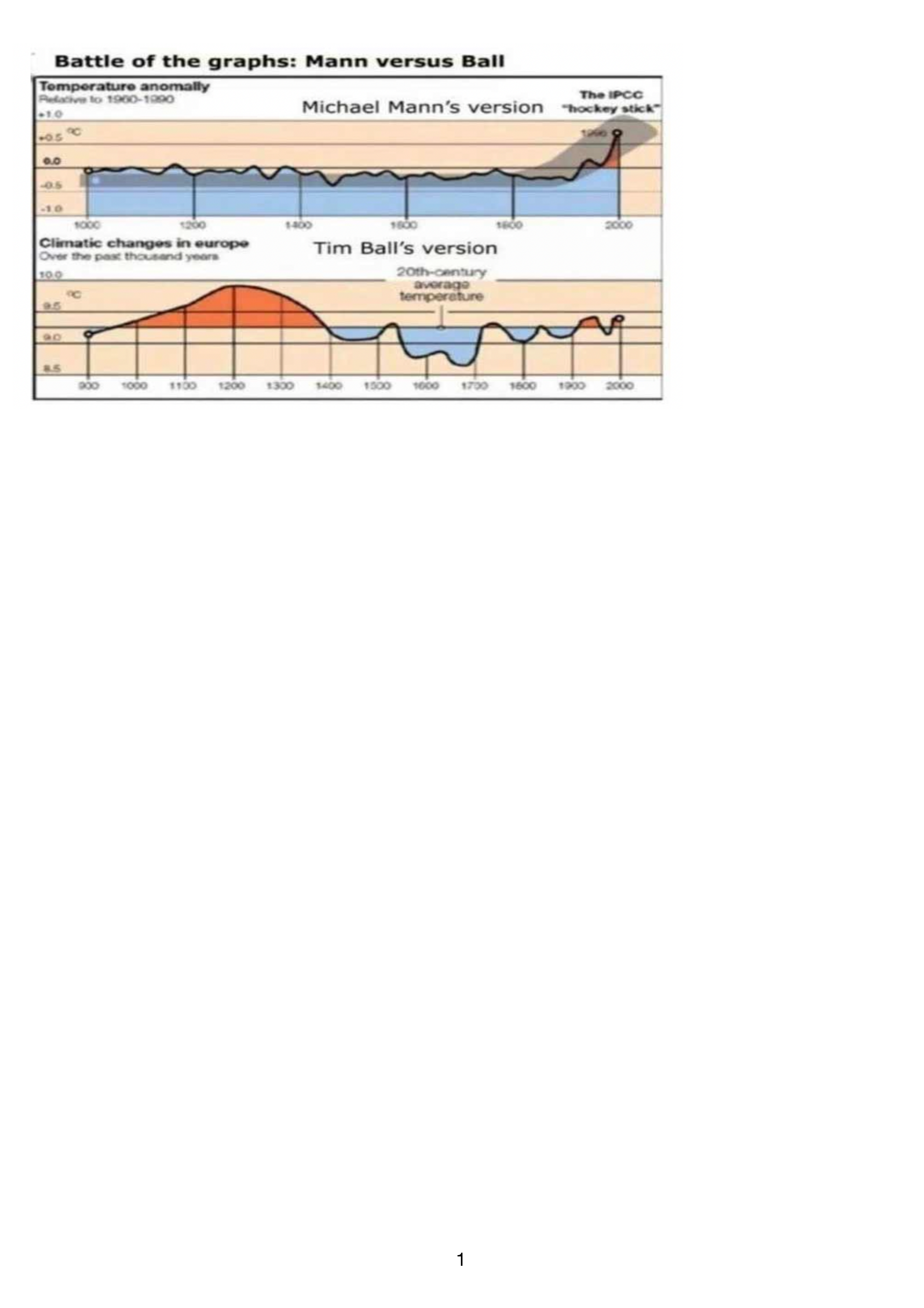}
\vspace{-26\baselineskip}
\caption{Two contradictory curves}
\label{fig: Mannball1}
\end{figure}

Let's not talk about the libel suit, brought by Mann against Ball and won by the latter, since Mann could not produce the evidence justifying his curve. Suffice it to point out that the Wegman report\footnote{Report prepared in 2006 by three statisticians – Edward J. Wegman (George Mason University), David W. Scott (Rice University) and Yasmin H. Said (The John Hopkins University) – at the request of Joe Barton, representative of the US House committee to validate climate policy.} – whatever criticisms it may have been subjected to – found that Mann's curve was grossly wrong, and based on significant statistical errors. One of the crucial points of the report (section 4) is that the method of Mann et al. creates a hockey stick shape even when fed random red noise, which the authors don't seem to notice. One can therefore dispute the climatological relevance of the curve in question. The fact is, for example, that the existence of the medieval optimum seems to contradict the idea that the temperatures currently observed would be without equivalent for a thousand years. Without evoking, like Ball, a "deliberate corruption" (see \cite{Bal}) of climate science, we can nevertheless underline the urgency of a more measured apprehension of global warming and of its consequences. Let us add, for good measure, that the Wegman report, in its conclusion (see \cite{Weg}, 10$^{th}$ point, 50), specifies that written by statisticians, it limited itself to examining the methodology of Mann and his collaborators, and that he focused on answering that question and not on whether or not the global climate was changing. "We have discussed paleoclimatology, write the authors, only insofar as it was necessary to clarify our discussion of statistical issues. The instrumented temperature record clearly shows that global temperatures have been increasing since 1850 CE. The way how this current era compares to previous eras is unclear due to uncertainties in the available records".
 
\subsection{The Keeling Curve}
 
 Let us examine now the question of the 50\% increase in CO$_{2}$ since the beginning of the industrial revolution. In appearance, it is clearly revealed on the following well-known diagram (see Fig. 2 - left). It shows the growth curve of CO$_{2}$ in the atmosphere since 1740. This curve – known as the Keeling curve\footnote{Named after Charles David Keeling, researcher commissioned by Roger Revelle, American oceanographer from the Kripps Oceanographic Institute, to direct an atmospheric carbon dioxide program in the mid-1950s. In July 1956, Revelle's team joined Keeling, who began continuous measurements of atmospheric carbon dioxide at the Mauna Loa Observatory, an observatory located on the Mauna Loa volcano in Hawaii. At the same time, measurements were carried out in parallel in Antarctica, Alaska and the Samoa Islands. Since that time, Revelle has made the study of the entire carbon cycle and the solubility of calcium carbonate a priority, accumulating data that is still used today by the IPCC and many researchers for studies of forecasting and climate modeling. The analysis of the observatory's atmospheric results is done in Hilo (see map, Fig. 2 - right)) where stratospheric balloons are also sent weekly from the old airport to assess the concentration of ozone and water vapor, while a site at Kulani Mauka collects rainwater, and a Lidar system measures air quality.} represents measurements made on ice cores to which we have added measurements of the Mauna Loa Observatory for the period 1958-1995. We obviously note a significant acceleration in the increase in the concentration of CO$_{2}$ since 1958 (see \cite{Mac}).
  
   \begin{figure}[h] 
	      \vspace{-1\baselineskip}
	   \includegraphics[width=8in]{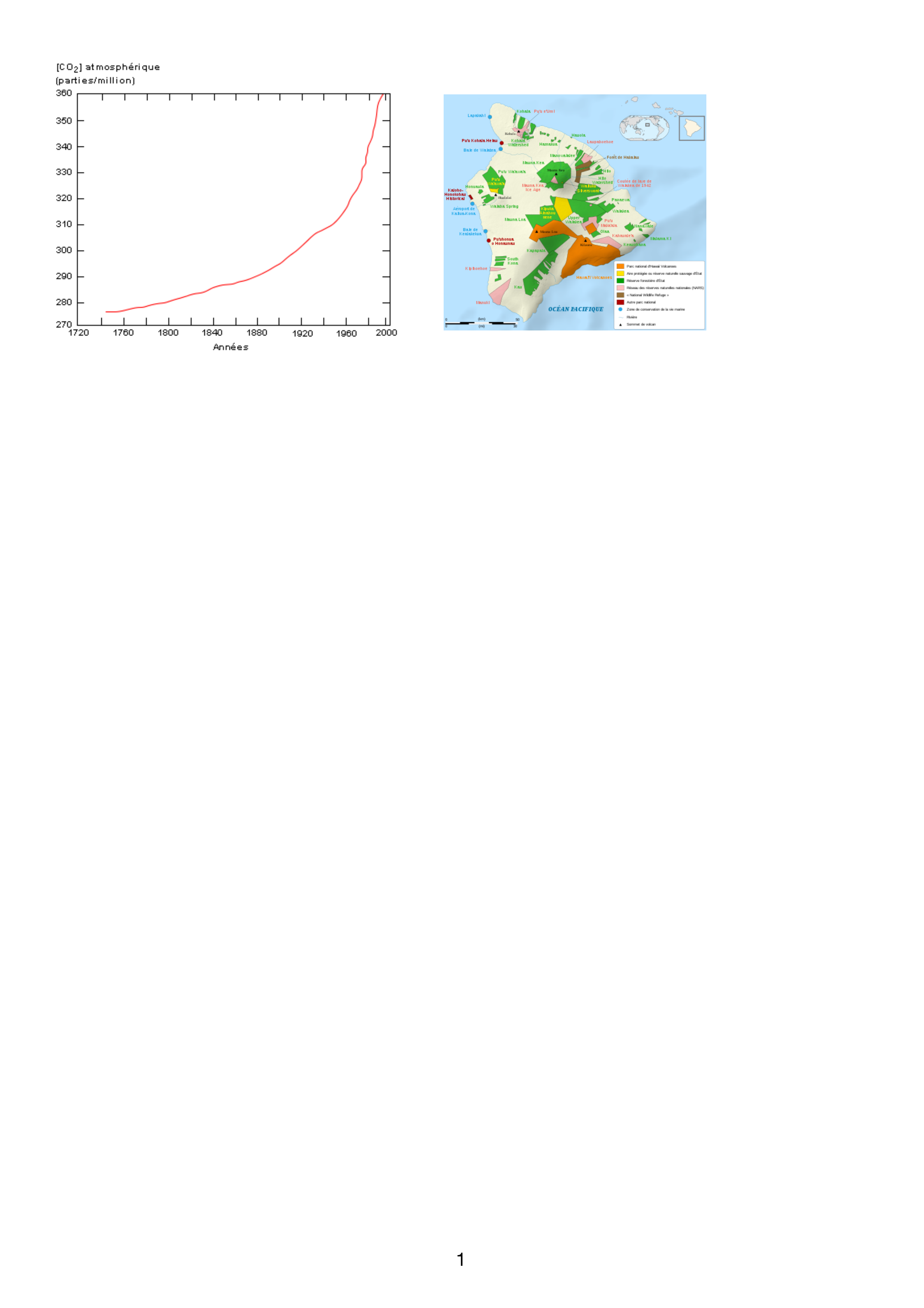} 
	      \vspace{-42\baselineskip}
	   \caption{Keeling's curve and map of Hawa\i}
	   \label{fig: Keel1}
	\end{figure}

	As can be seen, the amount of CO$_{2}$ in the atmosphere before the industrial revolution was around 280 ppm. It has therefore since increased by around 50\%, which cannot be without consequence on the temperature of the atmosphere in the vicinity of the Earth, and therefore on the climate. But the question, from there, is to know what meaning to give to this correlation. It would obviously be logical to consider that the increase in CO$_{2}$ is the {\it cause} of the rise in terrestrial temperatures. Unfortunately, it is rather the opposite that we observe. A graph by Fran\c ois Gervais (see \cite{Ger}) compares, for the period 1980-2005, the mean temperature $T(t)$ (solid line) on Earth at each instant $t$ (on an annual scale), to the annual variation $\theta(t)- \theta(t-1)$ of the rate $\theta(t)$ of CO$_{2}$, shifted by 6 months, i.e. $\theta(t+1/2)- \theta (t-1/2)$. With a suitable choice of scale for $T(.)$ and $\theta(.)$, we see that the curves overlap. We certainly deduce that the CO$_{2}$ level measured at times $t+1/2 = t+6$ months and $t-1/2 = t-6$ months seems linked to the temperature at time $t$. But, if this is the case, it is the temperature at time $t$ that influences the CO$_{2}$ level at time $t+1/2$, not the reverse (see Fig. 3).

\begin{figure}[h] 
\vspace{0\baselineskip}
\hspace{8\baselineskip}
  \includegraphics[width=3in]{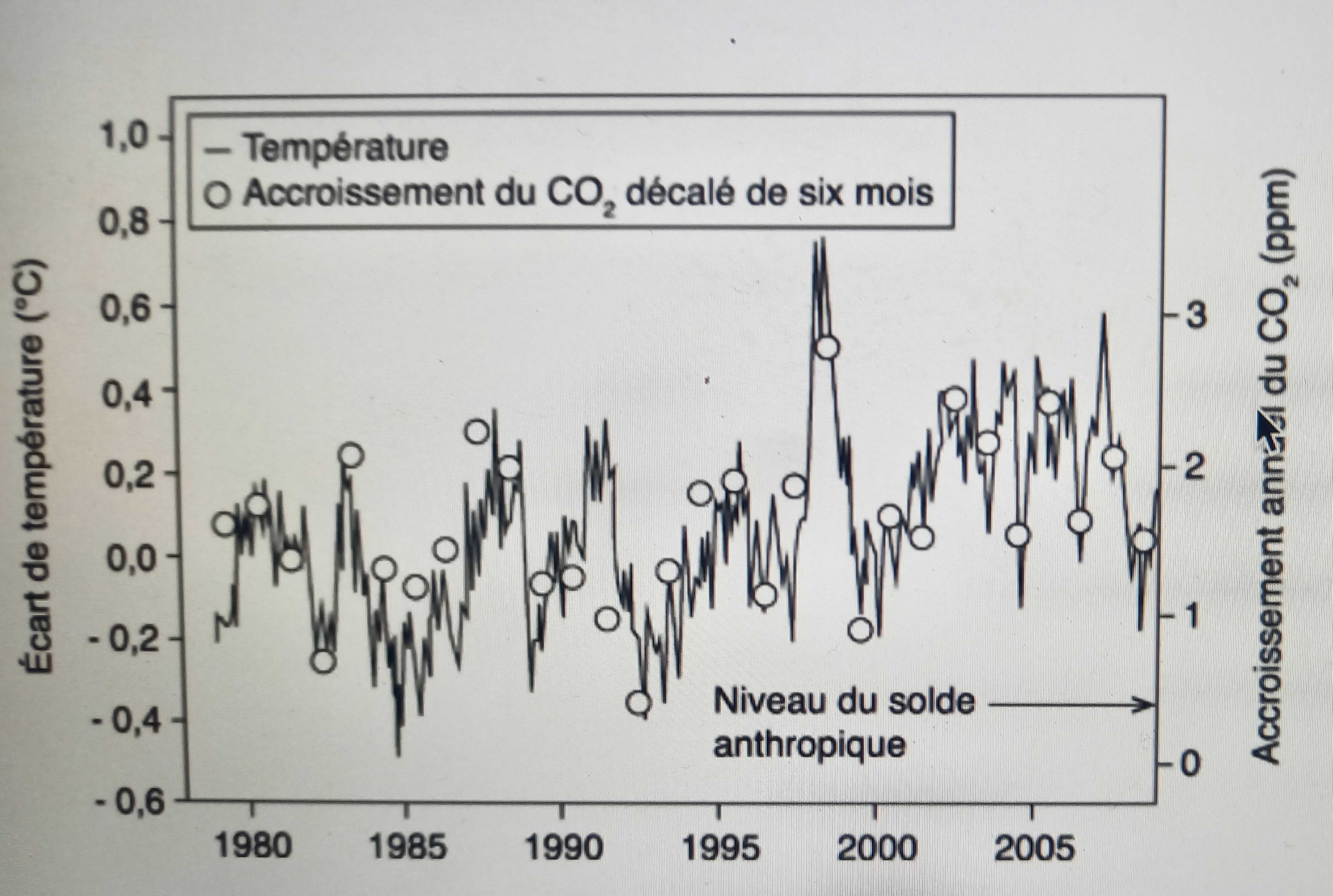}
\vspace{0\baselineskip}
\caption{Influence of temperature on CO$_{2}$}
\label{fig: CO21}
\end{figure}

Consequently, we cannot conclude from the increase in CO$_{2}$ that the temperature of the atmosphere in the vicinity of the Earth is rising, and neither, obviously, that there is a direct influence of this gas on the famous "greenhouse effect". The greenhouse effect of CO$_{2}$ exists, but it has been shown that it is saturated at the two main frequencies of the infrared spectrum of the molecule. If we observe the spectrum of a few tens of meters of atmosphere, in fact, it appears that this layer has become completely opaque at 20 and 70 terahertz (resp. 15 and 4.3 micrometers in wavelength), the two main infrared emission lines of CO$_{2}$. Heinz Hug (see \cite{Hug}) even showed that in the vicinity of the 15 micrometre band, doubling the carbon dioxide content would only induce a difference in residual transparency of 0.17\% over an altitude of ten meters, this band being "the most active in terms of the greenhouse effect, due to its proximity to the wavelength of the black body maximum", as François Gervais comments (see \cite{Ger}, 135). Translated into radiative forcing, this result leads, according to Hug, to a warming of 0.015$^\circ$ C, which is confirmed by other physico-chemical studies. Should we be alarmed for this?

\subsection{The greenhouse effect}

Known since the XIX$^{th}$ century, the greenhouse effect is linked to the following physical situation: the Sun emits radiation in visible light, ultraviolet and near infrared. The Earth reflects about 30\% of this radiation (this is called "albedo") and absorbs the rest. It itself emits radiation in the far infrared\footnote{The values are as follows: the greenhouse effect is measured by the difference between the infrared flux emitted by the surface of the Earth (on average 390 W.$m ^{-2}$), and the outflow at the top of the atmosphere (239 W.$m^{-2}$). The greenhouse effect is therefore 151 W.$m^{-2}$ (see \cite{Fou}).}. If there were no greenhouse effect, this last radiation would go back into space and it is estimated that the temperature on the surface of the Earth, instead of being on average around 14 at 15$^{\circ}$C, would rather be around -18$^{\circ}$ to -19$^{\circ}$C (-20$^{\circ}$C in some publications see \cite{Tho}). This is easy to prove.

\subsubsection{The temperature of the Earth without the greenhouse effect}

  Calculating the surface temperature $T_{S}$ of the Earth is quite easy to do from the following data:

Solar flux $\Phi_{0} = 1362\ W/m^2$;

Albedo $\alpha = 0.3$;

Emissivity/absorptivity of the atmosphere: coefficient $\epsilon = 0.76$;

Incident flow: $\frac{\Phi_{0}}{4}(1 - \alpha)$;

Stefan-Boltzmann constant: $\sigma = 5.670 367(13) \times 10^{-8}\ W.m^{-2}.K^{-4}$.

Stream sent: $\sigma T_{S}^4$;

Balance condition:
\[
\frac{\Phi_{0}}{4}(1 - \alpha) = \sigma T_{S}^4;
\]
From where:
\[
T_{S} = \frac{1}{\sigma}\left[\frac{\Phi_{0}(1 - \alpha)}{4}\right]^{\frac{1}{4}}.
\]

The numerical application here effectively gives a value of the order of 254$^\circ$ K, or -18$^\circ$ C.

It should be noted that the presence of an atmosphere associated with a greenhouse effect is likely to considerably modify the temperature at the surface of a planet, especially if this atmosphere is mainly formed of gas absorbing infrared radiation. Thus, the temperature of Mercury, a planet without an atmosphere, is around 167$^\circ$ C, while that of Venus, a planet whose atmosphere is made up of 96\% CO$_{2}$, is – as we have already said – 462$^\circ$ C.

\subsubsection{The temperature of the Earth with the greenhouse effect}
It must be taken into account, however, that part of the terrestrial radiation is returned to the Earth, due to cloud cover. The corresponding situation is shown in the following diagrams (see Fig. 4):

  \begin{figure}[h] 
  \vspace{0\baselineskip}
  \hspace{-2\baselineskip}
\includegraphics[width=8in]{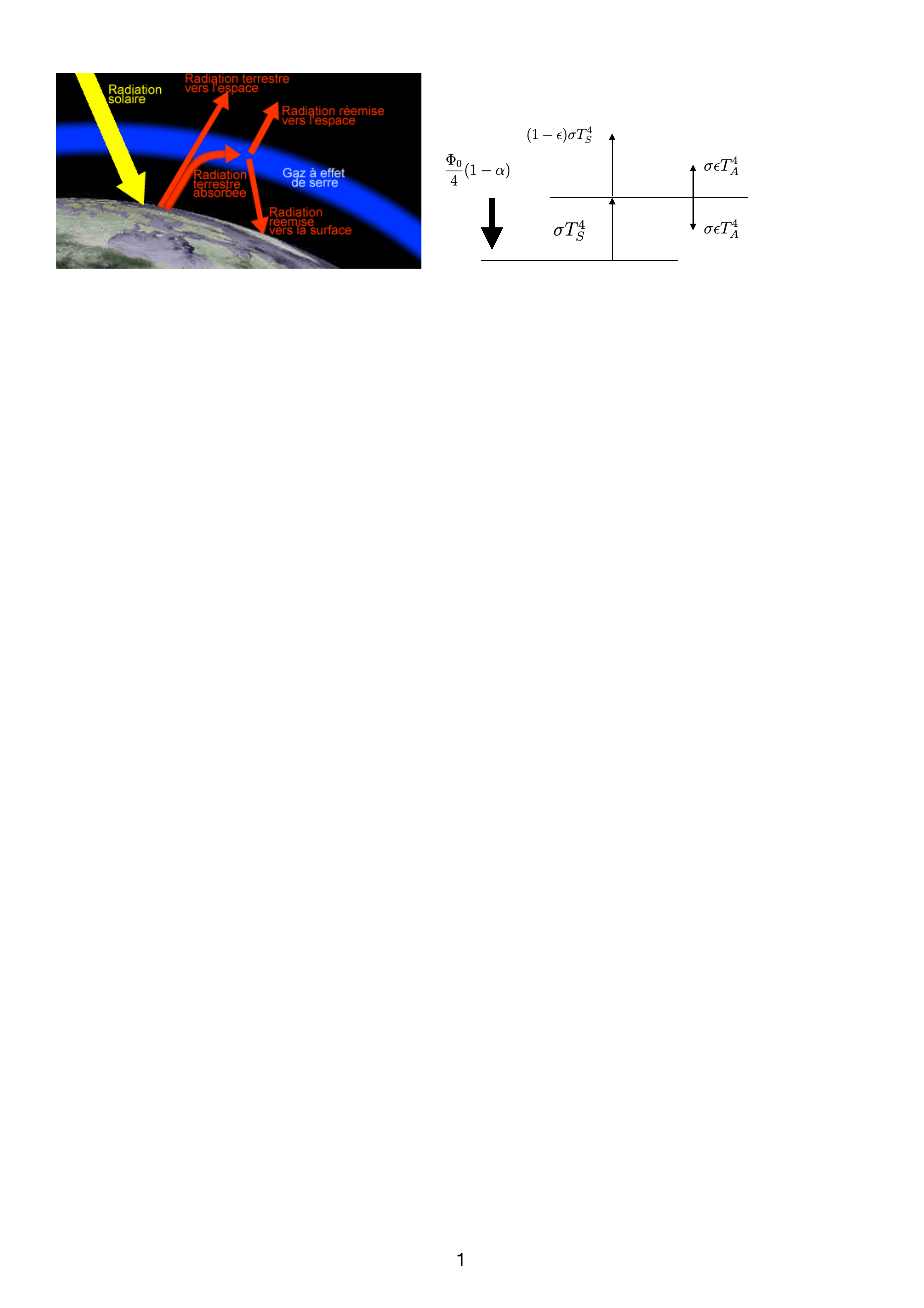}
\vspace{-44\baselineskip}
\caption{Earth radiation and solar flux}
\label{fig: Flux1}
\end{figure}

We always have the same solar flux $\Phi_{0}$, the albedo $\alpha$, the surface temperature of the earth $T_{S}$. Let $T_{A}$ be the temperature of the atmosphere.

The equilibrium condition of the atmosphere is given by:
\[
\sigma\epsilon T_{S}^4 = 2 \sigma \epsilon T_{A}^4, \quad \textnormal{hence:}\quad T_{A} = \frac{T_{S}}{2 ^{1/4}};
\]
The equilibrium condition of the earth's surface with the atmosphere is:
\[
\frac{\Phi_{0}}{4}(1 - \alpha) + \sigma \epsilon T_{A}^4 = \sigma T_{S }^4.
\]
or:
\[
\frac{\Phi_{0}}{4}(1 - \alpha) + \sigma \frac{\epsilon}{2} T_{S}^4 = \sigma T_{S }^4.
\]
Hence finally:
\[
T_{S} = \left[\frac{1}{\sigma} \frac{\Phi_{0}(1 - \alpha)}{4} \frac{1}{1 - \frac{1}{\ epsilon}}\right]^{\frac{1}{4}}.
\]

It is verified that the numerical application gives, in the latter case, an average terrestrial surface temperature of the order of 287$^\circ$ K, that is +14$^\circ$ C.

\section{Astronomical parameters and the theory of paleo-climates}

Before considering possible actions to limit the increase in the greenhouse effect – if indeed anthropogenic action modifies it in a significant way –, let us return to the past of the Earth as a planet of the solar system and examine the astronomical parameters of the climate explaining in particular the variations in temperature over time.

\subsection{The Milankovitch Theory}

The Serbian astronomer Milutin Milankovitch, after having studied the slow changes of the Earth's orbit, due to the interactions with the other planets of the solar system, was able to highlight three components of the orbital variability likely to generate changes of temperature on Earth and notable climatic variations (see \cite{Mil1}; \cite{Mil2};\cite{Mil3}). These are the famous "Milankovitch cycles", which concern:

– The eccentricity of the earth's orbit (period of 413,000 and 100,000 years);

–The inclination of the axis of rotation of the Earth (period of 41,000 years);

– The phenomenon of precession of the equinoxes (period of 23,000 and 19,000 years);

Milankovitch's theory was criticized for a long time, but its refinement and its extensions at the end of the 1970s (see \cite{Ber}) were able to show its validity. In particular, the periods that we have just listed could be reconstructed from a spectral processing of the signals (see Fig. 5) (see also \cite{Lan}).

  \begin{figure}[h] 
\vspace{-1\baselineskip}
  \hspace{5\baselineskip}
  \includegraphics[width=8in]{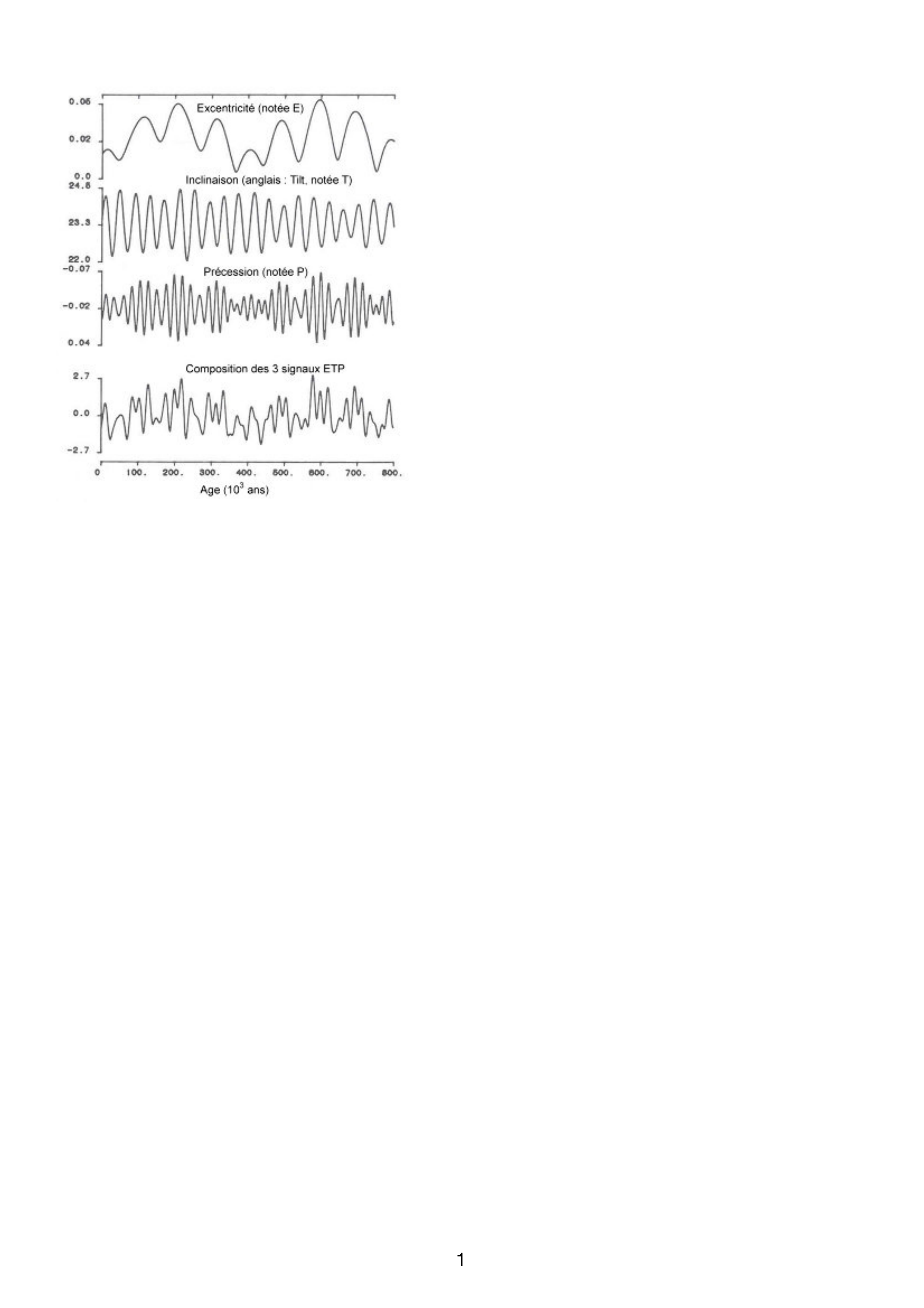}
\vspace{-36\baselineskip}
\caption{The Milankovitch Periodicity (from \cite{Cro})}
\label{fig: Milan1}
\end{figure}

More recent studies (see \cite{Hay}) have also made it possible to establish certain correlations between the variations of these orbital parameters and climatic changes on Earth, in particular the alternation of periods of glaciation and interglacial periods. Orbital changes, in fact, lead to large variations in the amount of sunlight received by the Earth during a given season (up to $\pm$15\%).

We can, moreover, reconstruct the variations in the volume of ice by using measurements of the isotopes of oxygen in the calcite (the "O" of CaCO3, for example) of the shells of foraminifera. Indeed, variations in $^{18}$O in seawater can be correlated to variations in ice volume\footnote{During the Ice Age, sea level was -130m. Consequently the $^{18}$O of the ocean was at a concentration of +1.5 per thousand higher than it is today. The measurement of $^{18}$O in the shells of foraminifera therefore makes it possible to reconstruct the variations in the volume of ice on the scale of millions of years.} and we also establish links between the variation in sea level and the variation in seawater isotopic composition and benthic foraminifera assays. This variability of the oxygen $^{18}$O is related to the variations of the direct radiation, in relation with the parameters of Milankovitch. The periodicity of Milankovitch and that of the glacial epochs are on the whole well correlated (see Fig. 6).

\begin{figure}[h] 
\vspace{0\baselineskip}
\hspace{3\baselineskip}
 	   \includegraphics[width=7in]{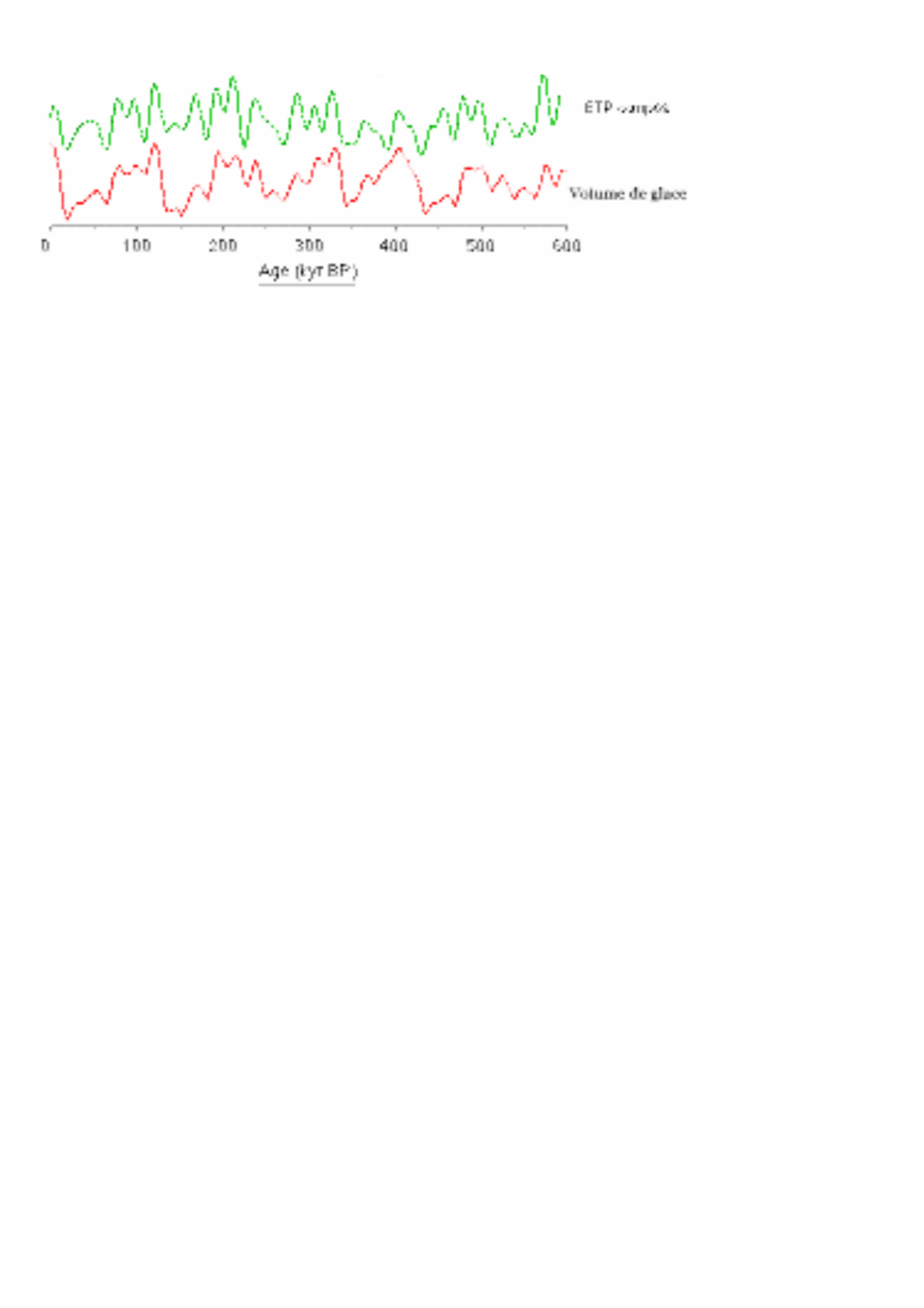} 
	      \vspace{-39\baselineskip}
	   \caption{Milankovitch periodicity and glaciations}
	   \label{fig: Donnees1}
	\end{figure}

It therefore appears that the periodic variations of the Earth's orbit are the stimulator of ice ages.

\subsection{Some problems of Milankovitch's theory}

Despite a good overall correlation with the facts, Milankovitch's theory is, in detail, far from perfectly coinciding with the experiment. Thus, all the cycles are not always present, and sometimes others whose origin is little known are added to the previous ones.

For example, the evolution of continental ice reveals a cycle of 100,000 years in the recent quaternary, mainly attributed today to eccentricity. However, this is in contradiction with the "historical" theory of Milankovitch who attributed the glacial variability to the summer insolation in the high latitudes, which in fact contains only a very weak contribution of the eccentricity.

The comparison between the volume of the glaciers and the summer insolation over the last 500 thousand years, in addition to the fact that it is not visually obvious, also reveals shorter cycles than the famous large cycles of 100,000 years (see Fig. 7).

\begin{figure}[h] 
\vspace{-1\baselineskip}
\hspace{5\baselineskip}
  \includegraphics[width=7in]{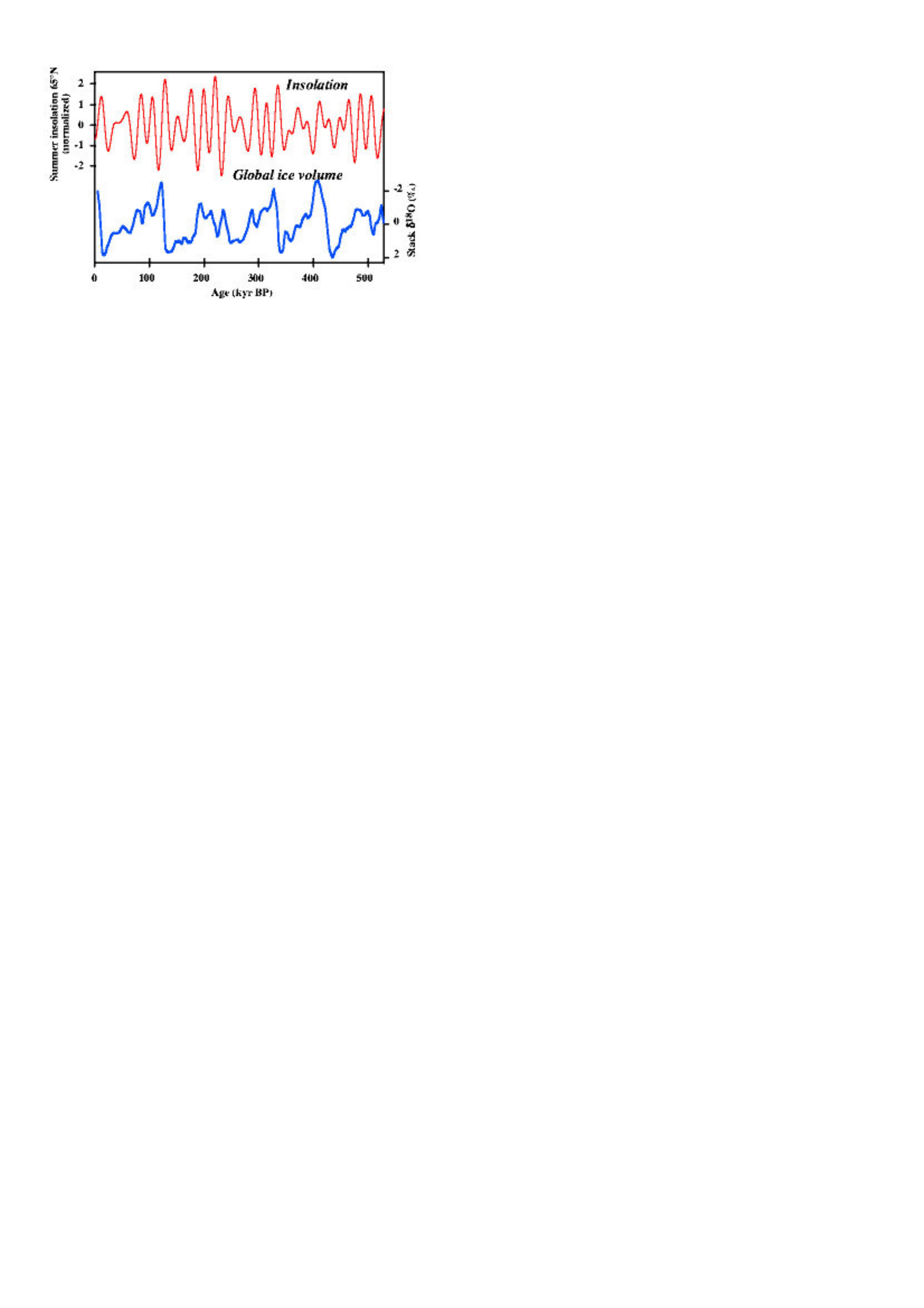}
\vspace{-39\baselineskip}
\caption{Volume variation of glaciers-summer insolation in high latitudes (65$^\circ$ N)}
\label{fig: Icecream1}
\end{figure}

To preserve the theory, we can then bring in certain non-linear mechanisms, but the matter is obviously a do-it-yourself affair. In terms of Popperian refutability, it would be better to recognize that summer insolation at high latitudes does not seem to play a direct role in the evolution of glaciers. And this, all the more so, since – if this were the case – the Antarctic ice should obviously evolve in the opposite way. However, this is not what we observe since drilling in Greenland and Antarctica has shown that the temperatures of the ice caps vary simultaneously. It thus seems necessary to formulate a more modern statement of Milankovitch's theory, in accordance with current observations.

\subsection{New Advancements}

Milankovitch's theory, however, is certainly worth saving. Independently of the question of paleoclimates and their variations, it had, among other things, the effect of making it possible to refine the geological time scale necessary for establishing a terrestrial chronology.

The use of a geological time scale is based primarily on sedimentary records collected around the world and linked together through significant events such as the disappearance or appearance of species or reversals of the Earth's magnetic field. All of this provides access to a relative timeline. Then, the recordings are dated absolutely using certain isotopic elements. The method is effective for the oldest records (about 100 million years), but astronomical dating is more accurate for the most recent sediments. This is where the Milankovitch cycles come in. The cycles observed in the paleoclimatic recordings are compared to them. The result is remarkable: the resulting absolute dating is accurate to within only 40,000 years over more than 10 million years. Most geological sedimentary records are thus now dated and calibrated using astronomical cycles.

An important difficulty concerning the use of Milankovitch's theory is to know if the cycles highlighted could have been affected by the deterministic chaos.

We know, in fact, that because of the chaotic behavior of planetary orbits – demonstrated by Jacques Laskar in 1989 (see \cite{Las}) – the uncertainty of calculations is multiplied by 10 every 10 million years, so that it is completely illusory to seek a precise solution concerning the position of the Earth beyond 100 million years. However, taking into account the current precision on the parameters of the solar system (position, mass of the planets, etc.), astronomical solutions can be used for dating sediments up to about 40 to 50 million years ago.

Beyond 100 million years, orbital variations probably also affected Earth's climates. Many Jurassic or Triassic sediments, for example, show marked cycles that are probably due to astronomical variations. However, there is still no way to compare them to simulations of the movement of the Earth at that time. One could naturally imagine that it is by studying these sedimentary cycles that it would be possible to find the constraints on the variations of the orbit and the orientation of the Earth at the time in question. But that would require sedimentary recordings of exceptional quality...

Even without the chaotic phenomena, other difficulties arise in reconstructing the astronomical past of the Earth, and therefore the ancient climates it may have experienced. We know, for example, that climatic precession and terrestrial obliquity are very sensitive to the history of the Earth-Moon system. Due to tidal effects, the Moon was closer to the Earth in the past and the Earth rotated faster. Few data, however, exist on this evolution and the value of the Milankovitch periods in the past presents for this a significant margin of uncertainty.

However, there might be a way around this problem. Some Milankovitch cycles are very stable over time and little affected by chaos. This is the case of the $\sim$400,000-year cycle of eccentricity, which essentially comes from gravitational disturbances generated by Jupiter and Saturn (giant planets very little affected by chaos over the long term). The search for this cycle in the records can be done over the whole Mesozoic era (up to 250 Million years ago) and some records seem to show such cycles. The future will tell if we can define a reliable geological time scale based on astronomy up to 250 million years back.

Another problem would be to know if there are Milankovitch cycles on other planets than the Earth. In principle, calculations of celestial mechanics and numerical integrations make it possible to calculate the evolution of the parameters of the orbit and the obliquity of the other planets over several million years. It should thus be possible to predict on which planet the climatic changes are expected to be significant. However, this remains highly speculative, and, from the point of view of our climate problems, less crucial for us.

\subsection{Milankovitch cycles and current era}

The central question is obviously to know exactly where we are in the Milankovitch cycles.

First of all, you should know that we are still discovering new ones today, of very long duration. Thus a recent study carried out by researchers from Toulouse (GET laboratory, CNRS-University of Toulouse 3) and Bremen (MARUM laboratory, University of Bremen) (see \cite{Mar}), shows that a long-term orbital factor temporized the seasonal dynamics of the climate between the Jurassic and the beginning of the Cretaceous. This study is based on an unprecedented analysis of the geochemical composition of belemnites, fossils of marine animals morphologically close to squid, having recorded the chemistry and temperature of seawater between -200 and -125 million years. Surprisingly, the results show a cyclical fluctuation in the carbon composition of the water every 9 million years. This is probably the longest cycle we have found so far.

Let us now try to answer the central question. According to the data we have, the last major deglaciation took place 128,000 years ago, at a time when the Earth in summer was relatively closer to the Sun and its axis of rotation was strongly inclined (24, 2$^\circ$). As we already reported in \cite{Par2}), a similar conjunction was revealed 11,000 years ago, when the Earth's climate experienced summers that were hotter than today's, a higher average temperature of about 2$^\circ$C and a sea level higher by about 2 meters. Currently, however, the Earth is farthest from the Sun in summer. As a result, the summers – apart from a few scorching days – should not be as hot as they could be and the winters should also in principle be rather mild. This can be related to the fact that during the last ten thousand years, the high latitudes of the northern hemisphere have gradually cooled: the permafrost initially restricted to very high latitudes has progressed towards the south, which seems to announce the establishment of a permanent snow cover at the high latitudes of the northern hemisphere, a premise for the next glaciation\footnote{However, recent measurements indicate a warming of the permafrost. During the decade 2007-2016, ground temperature near zero annual amplitude depth in the continuous permafrost zone increased by 0.39 ± 0.15°C. During the same period, discontinuous permafrost warmed by 0.20 ± 0.10°C. Permafrost in the mountains warmed by 0.19 ± 0.05°C and in Antarctica by 0.37 ± 0.10°C. Overall, the temperature of the permafrost would thus have increased by 0.29 ± 0.12°C. The observed trend would follow the arctic amplification of the air temperature increase in the northern hemisphere. In the discontinuous zone, however, ground warming occurred due to increased snow depth while air temperature remained statistically unchanged (see \cite{Bis}). But these local variations are not on a cosmo-geological scale}. However, the eccentricity of the Earth's orbit is currently particularly low, while the inclination of the Earth's axis of rotation is far from at its minimum. In such a configuration, the situation is not very favorable to the return of a glaciation, especially since, the Earth-Sun distance tending to decrease in the next thousand years, the situation will gradually evolve towards the establishment of increasingly hot summers in the northern hemisphere. Everything seems to indicate the absence of construction of an ice cap in the next centuries or millennia, and a natural warming of the summer periods. If so, the configuration would not be new. Although rare during the past hundreds of thousands of years (orbital situation of low eccentricity) it still occurred about 400,000 years ago (isotopic stage 11), when the interglacial was particularly long (about 30,000 years old). At that time, the 11,000-year pendulum played little, and the new 22,000-year cycle was established while sea levels were still high. At the time, there were wild vines in the Vercors and the sea level was higher (probably by ten meters) (see \cite{Mel}). In summary, the exit from our interglacial period is likely to be long and, even if, in the long term, we are going again towards the cold, not towards the hot, the latter will accompany us for a certain time, cumulating with the rest .

In any case, the warming that has been observed in recent decades is, in any case, too rapid to be linked to changes in the Earth's orbit, and too great to be caused by solar activity. Although many elements are still poorly understood in the Milankovitch cycles\footnote{The variation in eccentricity seems too slight to cause significant climate change, the cycles are not always easy to detect – especially at 65$^\circ$  north latitude (reference value because the land, which is preponderant there, reacts more easily to changes than the ocean), finally the predominant cycles are not always the same (the cycles of 100,000 years have been apparent for only about 1 million years, whereas before, cycles of obliquity (of 41,000 years) were dominant). We do not know what caused this transition – hence the call for non-linear effects.}, we cannot rely on them to explain the current situation.

\section{Provisional balance sheet}

Let's quickly summarize what we know.

1. The Earth's radiation balance is approximately 342 W.$m^{-2}$. Global warming represents roughly 1\% difference in relation to the radiative balance, or 342/100, that is to say about 3W.

2. There is therefore a "radiative forcing" of approximately 3 W.$m^{-2}$, which seems out of proportion with variations of the order of 0.5 W.$m^{-2 }$, attributed to fluctuations in solar power according to 11-year cycles. The other "astronomical" parameters also seem linked to an extremely long periodicity, far too long to be able to be involved in the phenomena that occurred during the hundred years that separates us from the beginnings of the industrial revolution.

3. Like the amount of CO$_{2}$, a greenhouse gas\footnote{For a gas to be "greenhouse", i.e. to be able to absorb infrared radiation, its molecule is composed of 3 atoms or of at least 2 different atoms. Thus carbon dioxide (CO$_{2}$), methane (CH$_{4}$), ozone (O$_{3}$), water vapor (H$_{2}$O) and CFCs are greenhouse gases, while oxygen (O$_{2}$), Nitrogen (N$_{2}$), Hydrogen (H$_{2}$), Argon (Ar), etc. are not.}, has increased by 50\% in the atmosphere since the industrial revolution, it is tempting to attribute the origin of "radiative forcing", and therefore of global warming, to our CO$_{2}$ emissions (which, however, represent only 0.04\% of the entire atmosphere).

4. Remember, however, that since 99\% of the atmosphere is composed of nitrogen and oxygen (gases that play no thermal role), the conservation of favorable planetary temperature conditions is based on only 0.43\% of the gases that make up the atmosphere. Among the contributors to the natural greenhouse effect, water is the main one in terms of concentration (H$_{2}$O: 0.39\%), followed far behind by carbon dioxide (CO$_{2}$: 0.039\%), methane (CH$_{4}$: 0.00018\%) and nitric oxide (N$_{2}$O: 0.000032\%). In absolute terms, water vapor contributes about 50\% to the greenhouse effect, clouds (therefore liquid water) 25\%, CO$_{2}$ 20\%, all other gases contributing about 5\%\footnote{These values are calculated for the greenhouse forcing, not the full radiation balance. But clouds, which reflect solar radiation, also absorb terrestrial infrared thermal radiation. The bottom line of both is that, for now at least, the clouds are cooling the planet slightly (see \cite{Myh}).}.
 
5. Given these latest data, we could say that it would be better to act on the main contributor to the greenhouse effect, namely water vapour. The problem is that we cannot permanently modify the composition of the water vapor in the atmosphere, to which, moreover, water vapor of anthropogenic origin contributes negligibly (see \cite{Bou}). This one, in fact, is in equilibrium with the oceans, so that if we try to increase the water vapor in the atmosphere, the oceans, in a very short time, will have absorbed all the excess. .

It is not forbidden, however, to be interested in the effect of the cloud cover of the Earth on its radiative budget.

\section{Clouds and Water Vapor}

Everyone agrees that clouds play a very important role in the climate system, not only because they cause precipitation but because they influence both solar radiation and infrared radiation emitted by Earth (see \cite{Har}). There is even reason to suppose that a fraction of the lower cloud layer practically controls the global temperature (see \cite{Kau}).

Clouds absorb little solar radiation, but they scatter it on all wavelengths (this is why we see them white) and can reach the reflectivity of fresh snow (80\%). Equally important is the influence of clouds on infrared thermal radiation. Except for the thinnest, they can be assimilated to black bodies, that is to say bodies absorbing all the infrared radiation received and emitting all that their temperature allows them. The clouds that have the greatest effect on radiation are the most extensive and persistent clouds, namely marine stratocumulus clouds and cirrus clouds, each covering about 20\% of the earth's surface. The first, which are low clouds (their base is between 500 and 1000 m) have a temperature not very different from the earth's surface, so that their contribution to the greenhouse effect is limited. On the other hand, their reflectivity is high and their effect on the albedo important. In contrast, cirrus clouds, the highest clouds encountered at the top of the troposphere, have very low temperatures (-60 to -70$^{\circ}$C), a low contribution to the albedo due to their transparency to solar radiation, but a great influence on the greenhouse effect.

It was only in the 1980s that it was possible, thanks to observations from meteorological satellites, to gather reliable information on
the extent and properties of the global cloud cover, which is permanently around 60\%.

  Although many points remain obscure\footnote{Clouds can of course warm or cool the Earth, depending on the ratio of the solar radiation they reflect to the amount by which they reduce the emission of that radiation to space. But, in warm regions of the tropical oceans, these solar and terrestrial radiation effects almost exactly cancel each other out, so tropical convective clouds do not appear to alter the Earth's energy balance as measured from space. However, we do not know why this balance occurs, or whether it is likely to change with global warming. The first simulations concerning the life cycle of these clouds and their radiative effect are fairly recent (see \cite{Har2}).}, we now have a fairly good idea of the radiative forcing of clouds, in other words the variation in the radiative balance of the Earth that would result from their total elimination. The effect is sometimes negative, sometimes positive, since it concerns both the albedo and the greenhouse effect.
 
  The figures are as follows: on an annual average, we find approximately - 47 W.$m^{-2}$ for the albedo effect and +29 W.$m^{-2}$ for the tight. Overall, therefore, the albedo effect outweighs the greenhouse effect and the resulting effect of clouds is to cool the planet. If, instantaneously, they became transparent, a simple subtraction shows that the radiation balance would increase by 18 W.$m^{-2}$, an amount already quite remarkable in itself. However, this balance resulting from two even larger quantities, it is certain that relatively small changes in cloud cover and/or its properties could have a considerable impact on the climate.

Note now that the radiative forcing corresponding to global warming being of the order of 3W$.m^{-2}$, this quantity represents approximately 1/6 of the radiative balance of the clouds, or 16\%. It would therefore suffice to modify it by 16\% to have an effect comparable to global warming.

Furthermore, the annual average net radiative forcing of clouds, which is estimated by comparing the reflected solar radiation and the outgoing solar radiation, shows that it is not uncommon to see, especially at mid-latitudes, clouds absorbing up to 70W.$m^{-2}$ of energy (see Fig. 8), which allows, since tropical clouds are neutral with respect to the Earth's energy balance, to draw some general quantitative conclusions.

	\begin{figure}[h] 
\vspace{0\baselineskip}
\hspace{5\baselineskip}
 	   \includegraphics[width=7in]{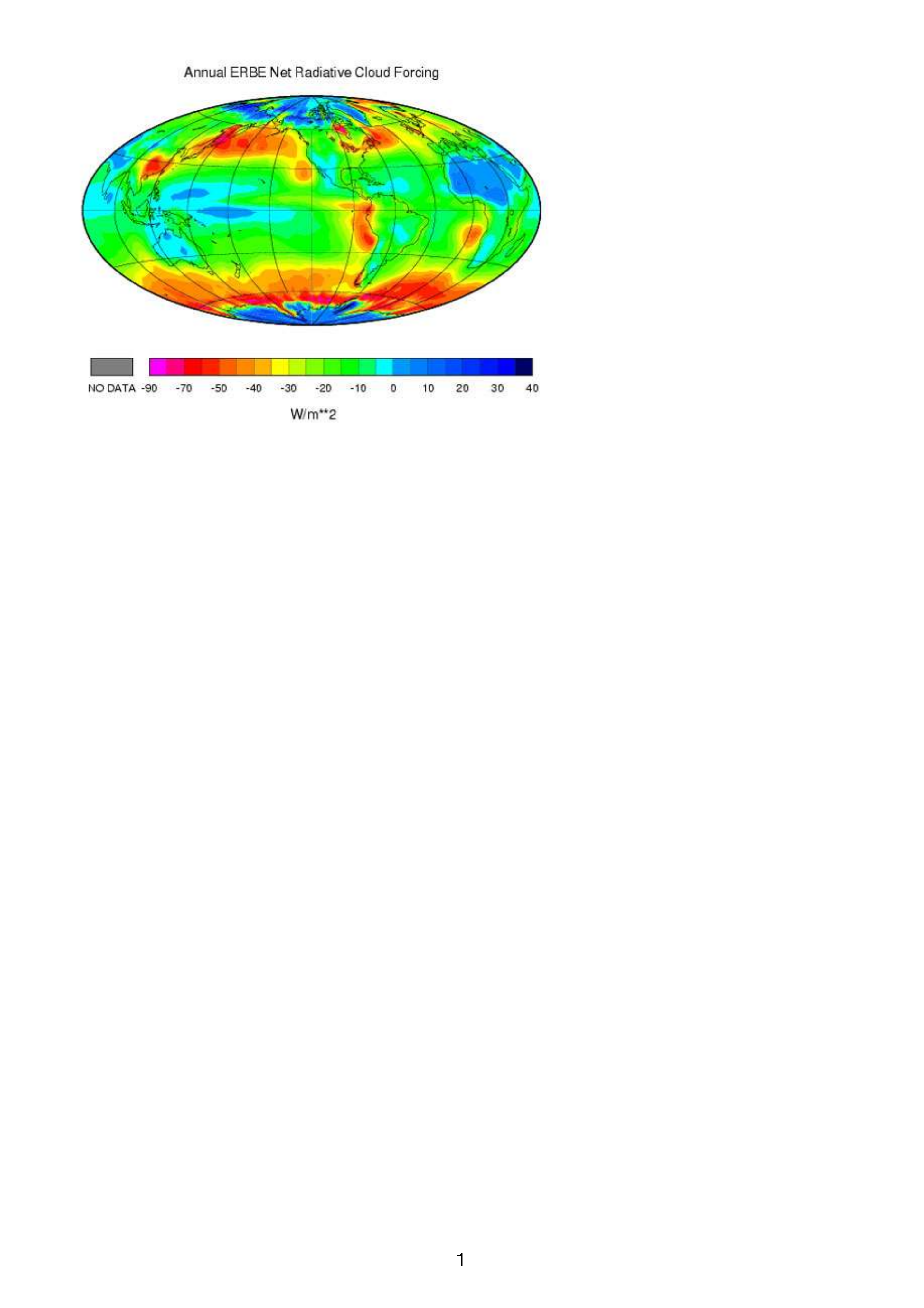} 
	      \vspace{-35\baselineskip}
	   \caption{	Average annual net radiative forcing of clouds (after D. Hartmann)}
	   \label{fig: Nuages1}
	\end{figure}
	
Indeed, we know that the radiative forcing is only 3W.$m^{-2}$, which represents in this case – if we retain this numerical value of 70W.$m^{-2}$ – 3 /70 of the amount subtracted, or 4.2\%. By extrapolating to the level of the whole Earth, we can therefore maintain that these 3/70 represent 4.2\% of the global clouds of the atmosphere. But since 70\% of the earth's surface is covered by oceans, the land surface represents only about 30\% of this surface. Consequently, clouds located strictly above the ground represent only 4.2\% of this 30\%. And so we can argue that the 3W increase in the Earth's radiation budget corresponds, roughly speaking, to the removal of $4.2 : 30$ = 14\% of purely terrestrial clouds. In other words, if we removed 14\% of the Earth's clouds, we would obtain an effect comparable to that of current global warming.

Furthermore, the following reasoning can be made:

A. The 1$^\circ$C increase in Earth's temperature, which corresponds to just over 3W.$m^{-2}$, accounts for about 7\% of the water-related greenhouse effect.
 
 B. In a cloudless tropical atmosphere, however, the greenhouse effect reaches 125 W.$m^{-2}$ (see \cite{Fou}) and, if the water vapor is removed, it is still 53W. $m^{-2}$, i.e. 35\% of the total value of the greenhouse effect (~151W.m$-2$ - see note 10). On the other hand, if we remove only the CO$_{2}$, it remains 94W.$m^{-2}$, that is, approximately, 75\% of the whole (94/125).

C. So, if the global temperature of the planet has risen by 1$^\circ$C and if we have, as we have said, a 7\% increase in the greenhouse effect, this fraction is not far from corresponding entirely to water vapor. We have in fact:
\[
\frac{94 \times 7}{100} \sim 6.6.
\]
Clearly, this result shows that an increase in the greenhouse effect is of the same order of magnitude as an increase in water vapor in an atmosphere where the CO$_{2}$ has already been removed.

\section{The Black Cloud}

For lack of being able to act on the flora and/or the water vapor, another non-decarbonizing solution against global warming would be to create an artificial protection reducing the quantity of solar energy received by the surface of the Earth. A few years ago, some scientists were able to imagine, for this purpose, devices of the "giant space screens" or "swarm of small satellites" type. The Lagrange point L1\footnote{A Lagrange point (denoted L1 to L5), or, more rarely, libration point, is a position in space where the gravity fields of two bodies in orbital motion around each other on the other, and substantial masses, provide exactly the centripetal force required for this point in space to simultaneously accompany the orbital motion of the two bodies. In the case where the two bodies are in circular orbit, these points represent the places where a third body, of negligible mass, would remain motionless with respect to the two others, in the sense that it would accompany at the same angular speed their rotation around their center of common gravity without its position in relation to them changing. The Lagrange point L1 of the Earth-Sun system, located between the Earth and the Sun – unstable like points L2 and L3 – is much closer to our planet. It should be noted that the presence of a screen at this location would undoubtedly prohibit the observations that are usually carried out there. Indeed, L1 is commonly used to measure solar activity (eruptions, cycles, solar wind) without this being affected by the Earth's magnetosphere. It is therefore the ideal position for space weather missions such as the one currently fulfilled by SoHO and ACE. The L1 point was also chosen to position the NEO Surveyor telescope intended to observe a large portion of space in which the near-Earth planets circulate.} of the Earth-Sun system could then appear as an ideal location for effectively shading the planet. Such an approach, however, presents several difficulties, such as maintaining the device in orbit against the pressure of solar radiation or, already, sending the required material - more than 10$^9$ kg - to the point L1, which is about a hundred times greater than anything man has sent into space to date.
	
	However, there is an alternative solution that would circumvent the problem: the use of micrometric dust grains as a heat shield.

A team from the University of Utah (see \cite{Bro}) recently proposed a scenario worthy of Fred Hoyle's novel, {\it The Black Cloud} (see \cite{Hoy} ). The researchers showed
that using dust to block out some sunlight would have the distinct benefit of mitigating climate change without having to long-term impact on our planet or its atmosphere. Their work is inspired by the process of planet formation, which is indeed accompanied by large amounts of dust, revolving around a host star. These rings of dust intercept the light of the star and re-emit it, including in the direction of the Earth, which also makes it possible to identify stellar systems with planets in formation. The idea would be to artificially create such a process by placing a small amount of dusty matter in a special orbit between the Earth and the Sun, which would block a large amount of sunlight with a very small amount of mass.

The overall effectiveness of such a shield obviously depends on its ability to stay in an orbit such that the Earth is shaded. So the team assessed the attenuation induced by different types of dust, then determined which orbits might hold the dust in position long enough to provide adequate shadow (see Fig. 9).
 
\begin{figure}[h] 
\vspace{-1\baselineskip}
\hspace{5\baselineskip}
  \includegraphics[width=5in]{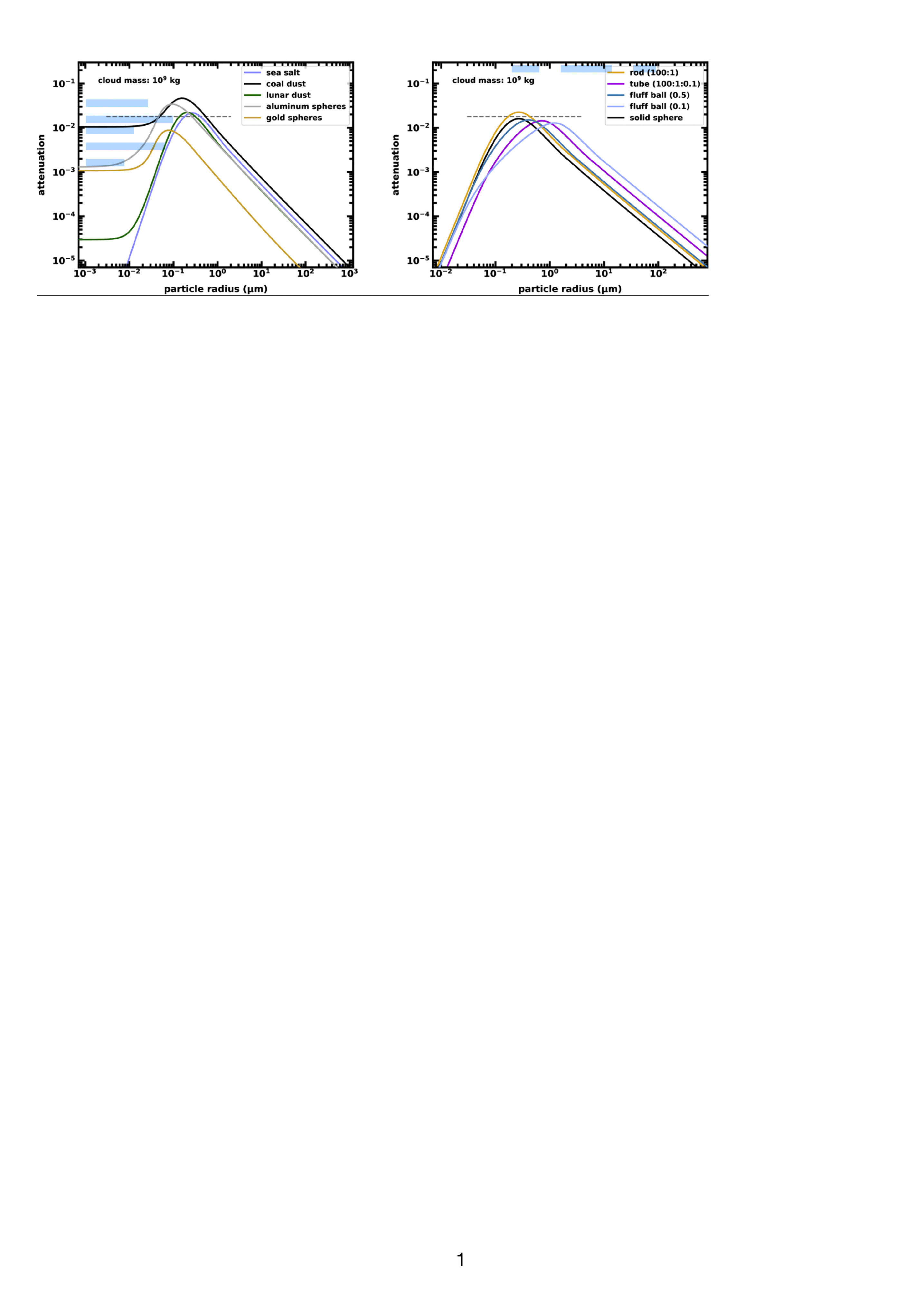}
\vspace{-28\baselineskip}
\caption{Attenuation of a monodisperse cloud of particles with a total mass of $10^9$ kg at point L1, as a function of radius: (left) attenuation for spherical particles of different types of materials; (right) attenuation for glass dust of different shapes. (According to \cite{Bro})}
\label{fig: Attenuation1}
\end{figure}

The calculations include variations in grain properties and orbital solutions as a function of lunar and planetary disturbances, said the researchers who were aiming for a reduction in solar irradiance of 1.8\%, corresponding to 6 days of attenuation per year.

The team estimates that around 10$^{10}$ kg of dust per year would be needed to get a meaningful climate result, depending on the properties of the dust and how the cloud is deployed. Many problems, however, arise. Computer simulations have indeed shown that the dust is easily deflected from its path by solar winds, radiation and gravity within the solar system. It would thus be necessary to provide a large reserve of dust, in order to be able to send it regularly after each dissipation of the cloud (see Fig. 10). Potential sources of this dust could be the Earth, the Moon or possibly a deflected asteroid.

	\begin{figure}[h] 
\vspace{1\baselineskip}
\hspace{9\baselineskip}
 	   \includegraphics[width=3.5in]{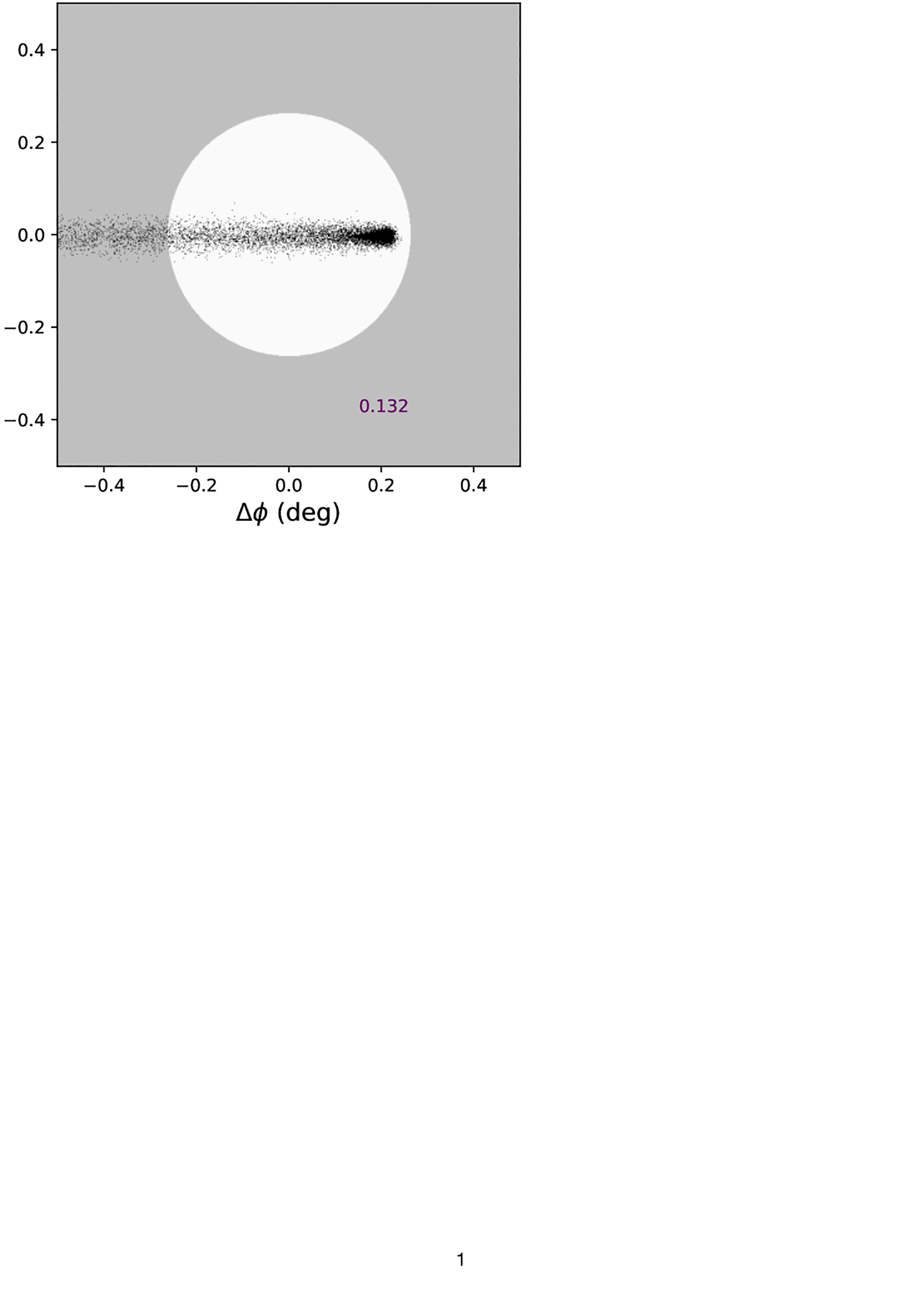} 
	      \vspace{-15\baselineskip}
	   \caption{Location of simulated micrometre-sized grains, launched continuously from an orbiter at L1, seen from Earth. The image shows the location of the grains 48 days after the start of the simulation (from \cite{Bro})}
	   \label{fig: Localisation1}
	\end{figure}

It appears from the study that one of the most promising approaches is to use fluffy grains with high porosity to increase the efficiency of extinction per unit mass, and to launch this material in directed jets from a platform located in orbit around L1, where the gravitational forces are balanced. Indeed, an object of negligible mass located on a Lagrange point remains immobile relative to the two bodies in orbital motion and rotates with them. Such a solution would, however, entail significant costs and effort.

A simpler and more economical approach would be to ballistically eject dust grains from the Moon's surface to the L1 point. The authors say that throwing lunar dust from the Moon could be a cheap way to shade Earth for several days. The advantages over a terrestrial launch lie in the existence of a ready-to-use reservoir of dust on the lunar surface and also in the lower kinetic energy required to reach a sunscreen orbit.

Note that the study has so far only explored the potential impact of the proposed strategy and has not determined whether these scenarios are logistically feasible.

The regular renewal of the dust could appear as a constraining aspect of this approach, but the temporary nature of the dust cloud is also an advantage: each cloud only persists for a few days before the dust is dispersed throughout the Solar System, there is therefore no risk that the Earth becomes cold and inhospitable\footnote{Another advantage would be that the cloud - which is not, like a screen or an orbital platform, a permanent opaque body - would only momentarily disturb the observations usually made from point L1.}.

\section{Conclusion}

1. The mere fact of increasing the temperature causes the greenhouse effect even if we do not form a cloud;

2. However, a decrease in water in the soil means that the saturation level (transformation of water into vapour) is not reached, which risks reducing cloud cover and increasing the greenhouse effect.

3. Plants create what is needed to enhance droplet nucleation. However, if there are fewer plants, the water will remain suspended in the air at the risk of evaporating or forming high altitude clouds, which, as we have seen above, contribute to the increase in the greenhouse effect.

4. It is therefore not certain that we will act most effectively on global warming through a policy of the "zero carbon" type, which it will be difficult to impose anyway, not only on developing countries, but also on industrialized countries, which, moreover, only reduce their carbon footprint by shifting it to the former and at the cost of social constraints leading to serious risks of destabilization. From this point of view, the “zero-carbon” policy in climatological matters is somewhat reminiscent of the “zero-Covid” policy adopted by China in epidemiological matters. Sooner or later this kind of policy becomes unbearable, and if we don't want everything to degenerate then, it is advisable to prepare an alternative policy well in advance.

5. It would be safer to act by preventing the terrestrial plant cover from deteriorating (because any disappearance of plants is problematic\footnote{Forests are also carbon sinks. Trees store CO$_{2}$ throughout their life. Deforestation therefore reduces the capacity of the global ecosystem to store CO$_{2}$ and contributes to increasing the greenhouse effect. However, the Earth's forest cover is steadily decreasing. Thus, the Amazon rainforest lost 72 million hectares between 1985 and 2018, that is to say 10\% of its extension in the space of 34 years, which cannot be without consequence on the climate. Generally speaking, according to a report by the IPBES (Intergovernmental Science-Policy Platform on Biodiversity and Ecosystem Services) of May 2019, the forest area observed on this date on Earth represents 68\% of that estimated in the pre-industrial era (\cite{Lec}, 494-495).}) and by managing the resources we have, even increasing them\footnote{Unlike the direct capture of CO$_{2}$ in the air and its storage by chemical processes converting the carbon into a solid form to bury it (energy-intensive and expensive process), the extension of these natural carbon sinks such as forests, peatlands or oceans, would be much more accessible. Of course, the land area needed to significantly reduce CO$_{2}$ levels by planting trees – which could be up to twice the size of India (see \cite{Bas}) – competes with other priorities, such as food crops. It could also have a negative influence on biodiversity and, with fires likely to increase with global warming, new forests would also be likely to quickly go up in smoke, causing the release of all the stored CO$_{2}$. Furthermore, we cannot expand the area of the oceans, which already absorb more than 30\% of humanity's carbon emissions. However, through "augmented weathering", a process of extracting and crushing rocks rich in minerals naturally absorbing CO$_{2}$, then spreading them on land or especially at sea, one could accelerate a process which normally takes place on scales of geological times of several tens of thousands of years. Of course, one question is whether this can be implemented on a sufficient scale, and at what cost. Another is to verify to what extent we could increase this absorption capacity of the oceans. It seems that a solution could be to artificially reinforce marine alkalinity or to "fertilize" them, that is to say to increase the density of phytoplankton which produce and sequester organic carbon by photosynthesis. But the secondary effects on ecosystems and the possibility of transposing this method on a large scale are still poorly known (On natural solutions to the problem of climate change, see \cite{Gris}; \cite{Dre}; \cite{Law}).}. The infiltration\footnote{The term unfortunately seems absent from many IPCC reports.} of the world's cultivated soils has been divided by 5 in a century due to poor agricultural practices and aberrant construction policies. However, without infiltration, floods are immediate in the event of major bad weather, drought is only deferred and direct sunshine is increased.

It is therefore not impossible, after all, that, contrary to intuition, the increase in the greenhouse effect is due – if not mainly, at least in part – to drought. It is commonplace to link the existence of deserts to the absence of water. It is less banal to realize that it is rather a lack of vegetation which potentially, if not causes global warming, at least contributes to it in a non-negligible way.

6. Finally, in the absence of being able to act on the flora or the natural clouds, one could envisage – as explained above – the creation of a heat shield around the Earth, in the form of a cloud of dust intended to shade temporarily – reversible solution and apparently without impact on the planet\footnote{It should nevertheless be verified that adding this dust to the already numerous debris (see \cite{Par4}) which revolves around the planet will have no consequences on the future launching of machines or the extra-vehicular exits of astronauts during the repairs they are required to make on the spacecraft they use.}.

\end{document}